# Smoothing analysis of HLSII storage ring magnets

WANG Wei(王巍),　HE Xiao-Ye(何晓业)*,　TANG Zheng(唐郑),　YAO Qiu-Yang(姚秋洋)

*University of Science and Technology of China, Hefei 230029, China*

**Abstract:** In order to improve the quality and stability of synchrotron light, Hefei Light Source has a major upgrade. Higher accuracy is necessary for installation and alignment of the storage ring magnets. It is not necessarily essential that the magnets are positioned exactly. In fact, the aim is to adjust neighboring magnets with a high accuracy to one another; in other words, these neighboring magnets are positioned on a smoothing curve. The paper presents an attempt to develop a reliable smoothing method based on curve fitting of least squares and iteration according to the structure characteristics of HLSⅡ. The method significantly reduces the adjusting amount and range of the storage ring magnets. It improves productivity by a factor of one times.

**Key words:** smoothing, HLSⅡ, curve fitting, least squares, iteration, relative error

**PACS**：06.06.Sx,29.90.+r,42.60.Jf

## 1 INTRODUCTION

Hefei Light Source（HLS）was designed and constructed in the 1980s, and formally opened to outside in 1992. The second stage of the project was constructed from 1999 to 2004[1]. In order to improve the quality and stability of synchrotron light, HLS has a major upgrade from June of 2010; the new storage ring of HLS is named HLSⅡ. Main parameters of HLSⅡ are better than HLS. The main characteristics of HLSⅡ in relation to metrology include complex structure with concentric assembly of parts, sub-millimeter accuracies required and high density of connected elements. It also means higher installation accuracy under the stringent demand of accelerator physics[2]. Many advanced methods are considered to meet this demand, with reference to the project experience of the other accelerators and synchrotrons of all over the world, smoothing analysis is considered as a fitness method to improve the accuracy and working efficiency. The smoothness of a storage ring refers to the quality of the relative positioning of a number of adjacent guiding components. According to the study result of CERN, as the major requirement for the geometry of an accelerator is that the relative errors must be very small (smaller than 0.1mm), an obligatory step for surveyors is to check the installation by measuring and-if needed-improving the smoothness of the machine alignment. Thus, the major requirement for the geometry of an accelerator is that relative errors must be as small as possible, in other words, the figure must be smoothing[3]. Shanghai Synchrotron Radiation Facility (SSRF) also considered the orbit smoothing during the installation project.

The paper presents an attempt to develop a reliable smoothing method based on the structure characteristics of HLSⅡ; the position algorithm of the storage ring magnets is given, and the real smoothing procedure is described together with the final alignment results of HLSⅡ.

## 2 SMOOTHING PROCEDURE

The storage ring of HLSⅡ consists of 77 magnets which are listed up in Table.1 and should be located in a regular octagon as shown in Fig.1 [4].

TABLE.1 Statistics of all magnets of the storage ring

| Types of magnets | Amount |
|---|---|
| Dipole | 8 |

Corresponding author, xyhe@ustc.edu.cn

| | |
|---|---|
| Quadrupole | 32 |
| Sextupole | 32 |
| Undulator | 5 |

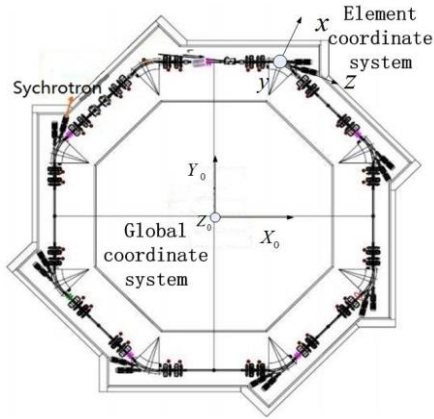

Fig.1 Layout of all magnets in HLSⅡ storage ring

The radius of curvature of dipole is 2.7729 m and the deflection angle of each dipole magnet is 45°[5]. A global coordinate system is built based on the geometrical symmetry and element coordinate system of each magnet is built according to their installation position (Fig.1). The fiducial of each magnet means equipped with internal reference marks defined in its own coordinate system; the center of its own coordinate system is magnetic field center as shown in Fig.2 [6].

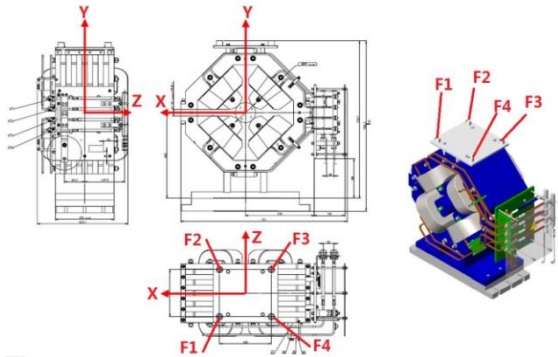

Fig.2 Internal reference marks and element coordinate system of each magnet

The process of smoothing includes six steps as shown in Fig.3

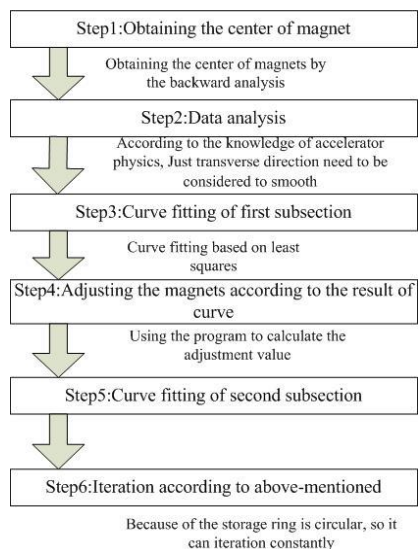

Fig.3 Smoothness procedure of HLSⅡ the storage ring

## 2.1 Center of magnets

In order to position the magnets to the storage ring, the installation control network is established by adding many points based on the first control network; then laser tracker is used to align these magnets according to the reference marks based on the installation control network.

After two times alignment, all the visible reference marks that located on the outer skin of the magnets are measured. By using the measured values of the reference marks, the magnetic field center of magnets can be indirectly obtained by the backward analysis; the backward analysis is based on the coordinate transformation [7]. Because of the actual and theoretical position of reference marks do not match completely; the coordinate transformation is also based on the least squares, the Spatial Analyzer(SA) software can be used to process the least squares transformation by the best-fit function.

By comparing the actual positions and theoretical positions, all magnets are positioned in a little error. In fact, magnets are positioned around an unknown mean trend one among an infinity contained within the envelope of maximum errors. The installation errors' statistical nature is essentially Gaussian: the aligned elements are randomly and normally distributed around this mean trend curve as shown in Fig.4 [8].

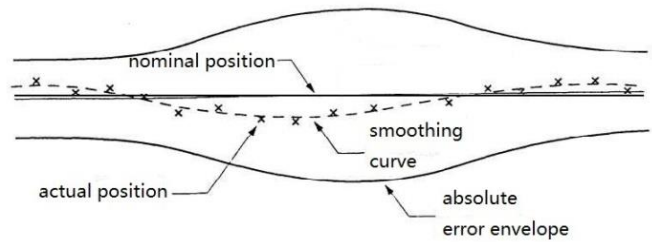

Fig.4 Position of magnets with respect to theoretical orbit

In this paper, there is a new opinion that the error of every magnet should be displayed in its element coordinate system not in global coordinate system because of the shape of HLSⅡ is a regular octagon as shown in Fig.1, there is an angle between the axis of global network coordinate system and beam line direction. The one-dimensional error of beam transverse motion will has two components in x axis and y axis of global network coordinate system. It is not beneficial to track the change of beam motion immediately.

## 2.2 Analysis

After obtaining the error data of all magnets, data analysis is necessary. According to the knowledge of accelerator physics, the orbit of beam includes longitudinal direction (Z axis) and transverse, the transverse also includes horizontal (X axis) and vertical (Y axis), as shown in Fig.1.

According to the closed orbit distortion formula.1 and the simulation code MAD that developed by CERN, the effective of alignment error of the all kinds of magnets can be studied [11]. The study shows that the most significant closed orbit distortion is caused by the error of transverse of quadrupole and sextupole and the error of rotation of dipole [6, 12].

$$(u_0^2(s))^{1/2} = \begin{cases} \dfrac{\beta(s)^{1/2}}{2\sqrt{2}|\sin \pi \upsilon|} \Delta u_{q.rms} \sqrt{\Sigma \beta(KL_q)^2} \\ \dfrac{\beta_s^{1/2}}{2\sqrt{2}|\sin \pi \upsilon|} [(\Delta B/B)_{b.rms} \, or \, \Delta \phi_{b.rms}] \sqrt{\Sigma \beta(L_b/\rho)^2} \end{cases} \quad (1)$$

In formula (1), $u_0(s)$ is the value of closed orbit distortion, $\beta(s)$ is the function value of $\beta$ in the observation point, $K$ is the strength of quadrupole, $L_q$ is the length of quadrupole, $L_b$ is the length of dipole, $\beta$ is the $\beta$ function of error point, $\rho$ is the bending radius of beamline, $\Delta u_{q.rms}$ is the rms of all quadrapoles, $\Delta \phi_{b.rms}$ is the rms of all dipoles.

The rotation of dipole has a big influence for the closed orbit distortion according to the context, and some methods have been taken to decrease the rotation error of dipole as shown Fig.5. In fiducical of dipole, the level and clamp were used to level the under polar of dipole, so the above polar of dipole was also seen as a level ignoring the manufacturing error. At the same time, a spirit level was placed on the top of the dipole and marked the tilt value. During the installation process, the laser tracker was used to adjust the dipole, then the spirit level was used to check the tilt of dipole. If the tilt value was not match with the marked value, adjusted the dipole according to the tilt value. So the rotation error of dipole is so small that to be ignored.

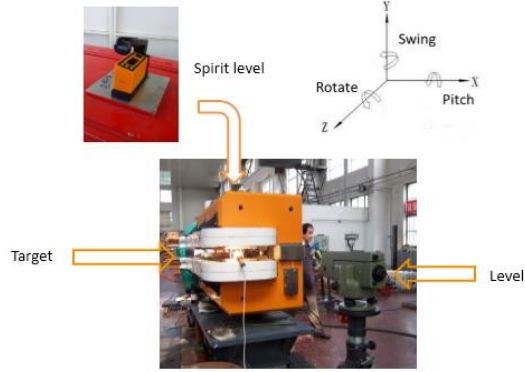

Fig.5 Adjusting the rotation of dipole by spirit level

Because of the error of longitudinal direction (Z axis) does not have a big influence for the closed orbit distortion, it is not considered by smoothing process and just transverse directions (X axis and Y axis) need to be considered. In order to simplify the smoothing process, all magnets can be considered as a point and their length and rotation are ignored.

**2.3 Smoothing based on the polynomial of least square**

Which magnets need to be adjusted according to the traditional method? Mean square error method is often used to judge whether these magnets should be adjusted.

$$\delta = \sqrt{\dfrac{\sum_{i=1}^{n} a_i^2}{n}} \quad (2)$$

In this formula, $\delta$ is RMS and $a_i$ is the difference of these magnets' actual position and theoretical position [13]. In addition, it is required that each single deviation is smaller than 2δ or 3δ. The storage ring of HLSⅡ consists of 77 magnets, and calculated by the formula, δ=0.082mm, this magnets need to be adjusted if their deviation are bigger than 2δ. According to the statistics, 17 magnets need to be adjusted and account for 22% of total magnets, the average adjusting value of the 17 magnets is 0.12mm as shown in Fig.6.

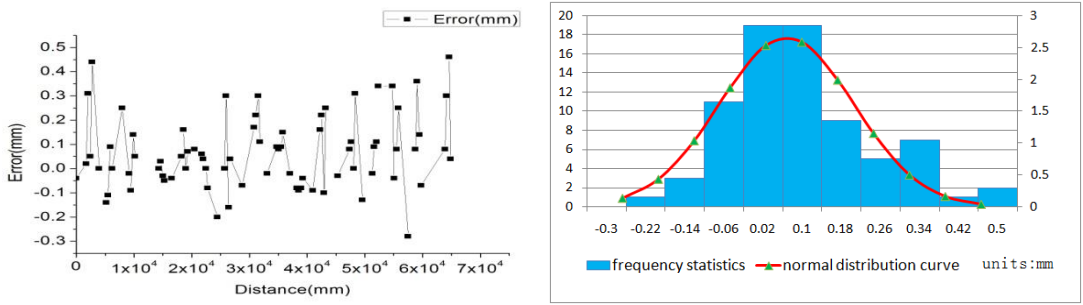

Fig.6 X axis error distribution of mathematical statistical based on theoretical orbit

If the magnets are adjusted on the basis of traditional way, the adjusting amount and range of the magnets are big and not beneficial for the smooth implementation of project. As we know, for the running of an accelerator, it is not necessarily essential that the magnets are positioned exactly. On the contrary, the aim is to adjust neighboring magnets with a high accuracy to one another. If neighboring magnets are positioned on a smoothing curve, systematic deviations between the nominal and the actual positioned will be unimportant. In order to make sure the neighboring magnets are positioned in a high relative accuracy, considering the structure characteristics of HLSⅡ storage ring, the method that the curve fitting of least squares of subsection and iteration of smoothing is developed to reduce the adjusting amount and range of the storage ring magnets.

The mathematical treatment of least squares is that observation functions $y_i = f(x_i)$ and function space $\Phi = span\{\varphi_0, \varphi_1 \cdots \varphi_m\}$; obtaining the function $\varphi^*$ and

$$\|f - \varphi^*\|_2 = \min_{\varphi \in \Phi} \|f - \varphi\|_2 \qquad (3)$$

Cause

$$\varphi(x) = \sum_{j=0}^{m} a_j \varphi_j(x), \varphi^*(x) = \sum_{j=0}^{m} a_j^* \varphi_j(x) \qquad (4)$$

So

$$S(a_0, a_1 \cdots a_m) = \|f - \varphi\|_2^2 = \sum_{i=1}^{n}\left[ y_i - \sum_{j=0}^{m} a_j \varphi_j(x_i)^2 \right] \qquad (5)$$

So the question is equal to

$$S(a_0^*, a_1^* \cdots a_m^*) = \min_{a_0, a_1, \cdots a_m \in R} S(a_0, a_1, \cdots a_m) \qquad (6)$$

For the necessary condition

$$\frac{\partial S}{\partial a_k} = 0, \quad k=0,1 \cdots m \qquad (7)$$

Obtain

$$\sum_{j=0}^{m}(\varphi_j,\varphi_k)a_j=(f,\varphi_k),\quad k=0,1,\cdots m \quad (8)$$

Matrix form

$$\begin{pmatrix}(\varphi_0,\varphi_0)&(\varphi_0,\varphi_1)&\cdots&(\varphi_0,\varphi_m)\\(\varphi_1,\varphi_0)&(\varphi_1,\varphi_1)&\cdots&(\varphi_1,\varphi_m)\\\vdots&\vdots&\cdots&\vdots\\(\varphi_m,\varphi_0)&(\varphi_m,\varphi_1)&\cdots&(\varphi_m,\varphi_m)\end{pmatrix}\begin{pmatrix}a_0\\a_1\\\vdots\\a_m\end{pmatrix}=\begin{pmatrix}(f,\varphi_0)\\(f,\varphi_1)\\\vdots\\(f,\varphi_m)\end{pmatrix} \quad (9)$$

According to this mathematical functions, the code of curve-fit can be coded by the MATLAB program [14].

In this context, the center of every magnet is obtained. The quadrupole located in the inject position of the storage ring is considered as the start, and then all magnets are located according to the relative position relationship of each other [15, 16]. Firstly, X axis is smoothed, the first sector of octagon is considered as the first subsection, and then the first side of octagon is considered as the second subsection, the two subsections have half common points as shown in Fig.7; by that analogy of the other magnets. When the first sector's smooth curve is fitted by least squares, δ can be obtained according to it. Adjusting the errors and then fitting the second smooth curve; iterating the other subsections according to this method [17].

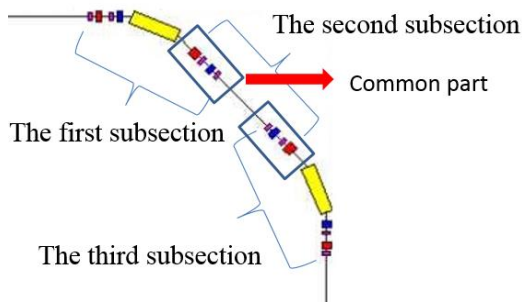
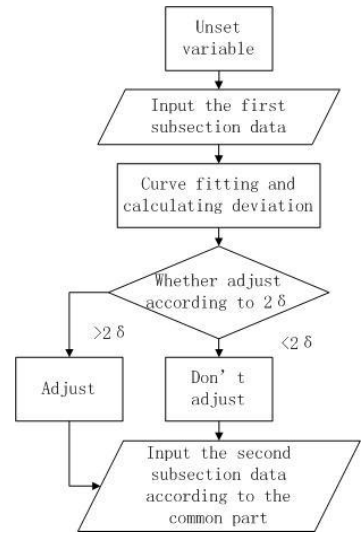

Fig.7 Iteration curve fitting of every subsection     Fig.8 Flow diagram of data processing

All the iteration steps of best-fit can be executed by MATLAB program as shown in Fig.8, and the relative figures also can be obtained through the program, just like the next pictures.

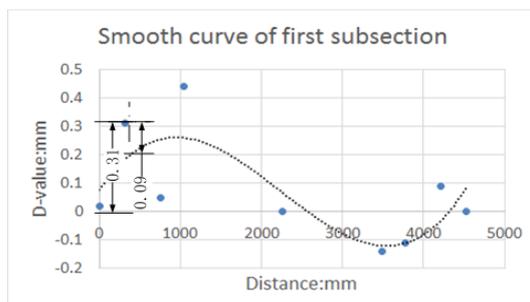
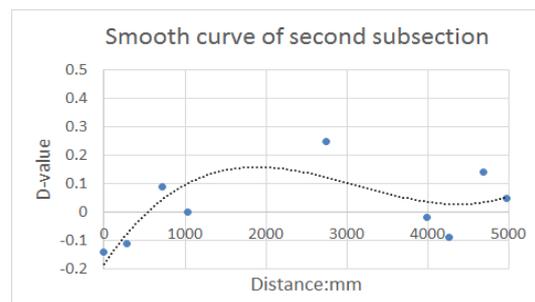

Fig.9 First subsection curve fitting　　　　　　Fig.10 Second subsection curve fitting

As shown in the Fig.9, the absolute error of the second point is 0.31mm, should be adjusted according to the traditional requirement; but the relative error is 0.09mm relative to the smoothing curve, should not be adjusted.

## 3 RESULT

After the smoothing process, only 10 magnets need to be adjusted and the average adjusting value of 10 magnets is less than 0.05mm (Fig.11), the method greatly reduces the adjusting amount and the average adjusting value. Y axis is also processed with the same method at the same time. Because of level and spirit level were used to help adjust the Y axis of all magnets, just only three magnets' Y axis need to be adjusted for smoothing[12,18]. The smoothing result of Y axis as shown in Fig.12.

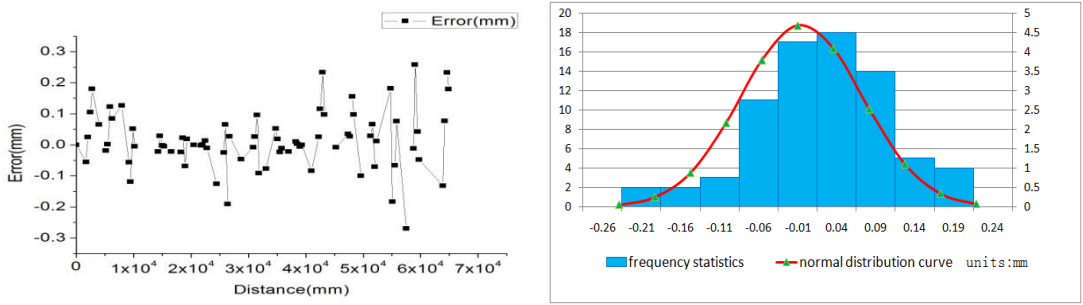

Fig.11 X axis error distribution of mathematical statistical based on smoothing orbit

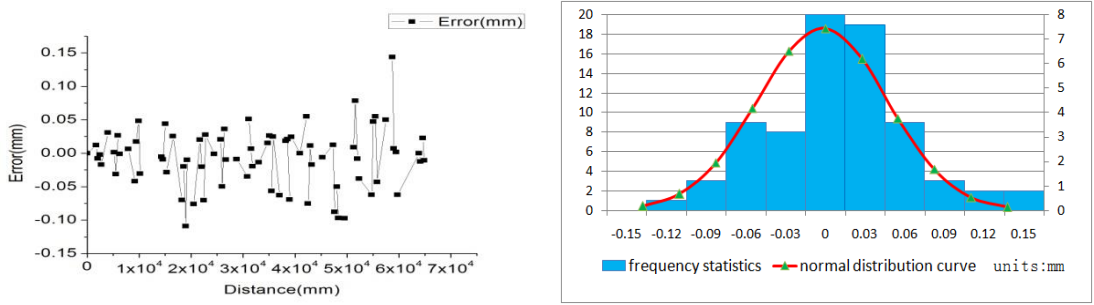

Fig.12 Y axis error distribution of mathematical statistical based on smoothing orbit

The perimeter of the storage ring can be roughly obtained through estimating the center point of magnets, the mechanical perimeter of HLS Ⅱ storage ring is about 66.171236m; the actual perimeter that calculated by accelerator physics formula is:

$$C = \frac{c}{204.03Mhz/45} = 66.17935 \ m \tag{10}$$

In the formula 10, the C displays the perimeter; the c displays the light speed. The mechanical perimeter and the actual perimeter are similar, so it also proves the result is right [19].

## 4 CONCLUSION

A smoothing criterion has been developed which reflected beam alignment tolerances and still allows the number of magnet adjustments to be minimized. The technique maybe still needs improvement to make it a general smoothing method, but its application to the HLS Ⅱ has already been highly successful. At the January of 2014, the storage ring was successful to encircle.